\documentclass[12pt]{article}
\usepackage{mathrsfs}
\makeatletter \@addtoreset{equation}{section} \makeatother
\renewcommand{\theequation}{\thesection.\arabic{equation}}
\newcommand{\ba}{\begin{array}}
\newcommand{\ea}{\end{array}}
\newcommand{\beq}{\begin{equation}}
\newcommand{\eeq}{\end{equation}}
\newcommand{\bea}{\begin{eqnarray}}
\newcommand{\eea}{\end{eqnarray}}
\def\bce{\begin{center}}
\def\ece{\end{center}}

\def\nonu{\nonumber}

\def\pa{\partial}
\def\al{\alpha}
\def\be{\beta}

\def\eps6{{\displaystyle \mathop{\epsilon}^{6}}{}}
\def\g6{{\displaystyle \mathop{g}^{6}}{}}

\def\nab6{{\displaystyle \mathop{\nabla}^{6}}{}}
\def\0{{\sst{(0)}}}
\def\1{{\sst{(1)}}}
\def\2{{\sst{(2)}}}
\def\3{{\sst{(3)}}}
\def\4{{\sst{(4)}}}
\def\5{{\sst{(5)}}}
\def\6{{\sst{(6)}}}
\def\7{{\sst{(7)}}}
\def\8{{\sst{(8)}}}

\def\ba{\begin{array}}
\def\ea{\end{array}}
\def\beq{\begin{equation}}
\def\eeq{\end{equation}}
\def\be{\begin{equation}}
\def\ee{\end{equation}}

\def\eps{\epsilon}

\def\ba{\begin{array}}
\def\ea{\end{array}}
\def\beq{\begin{equation}}
\def\eeq{\end{equation}}
\def\be{\begin{equation}}
\def\ee{\end{equation}}

\def\eps{\epsilon}

\def\eps6{{\displaystyle \mathop{\epsilon}^{6}}{}}

\def\nab6{{\displaystyle \mathop{\nabla}^{6}}{}}

\newcommand{\bean}{\begin{eqnarray*}}
\newcommand{\eean}{\end{eqnarray*}}

\begin{document}
\thispagestyle{empty} \addtocounter{page}{-1}
   \begin{flushright}
%PUPT-2395 \\
%CALT-68-nnnn \\
%{\tt hep-th/yymmnnn}\\
\end{flushright}

\vspace*{1.3cm}
  
\centerline{ \large \bf
 Fermionic   Construction in the Supersymmetric Coset Model
}
%\vspace*{0.3cm}
%\centerline{ \Large \bf  
% } 
%and }
\vspace*{1.5cm}
\centerline{{\bf  Changhyun Ahn$^\dagger$},
  {\bf  Jaesung Hong$^\dagger$},
  and {\bf Man Hea Kim$^{\ast}$
    %\footnote
    %{Address after April 1, 2020:
% Asia Pacific Center for Theoretical Physics, Pohang 37673, Korea}$}
%\footnote{On leave from the Department of Physics, Kyungpook National University, Taegu
%  702-701, Korea and 
%address until Aug. 31, 2011:
%Department of Physics, Princeton University, Jadwin Hall, 
%Princeton, NJ 08544, USA}
}} 
\vspace*{1.0cm} 
\centerline{\it 
$\dagger$ Department of Physics, Kyungpook National University, Taegu
41566, Korea} 
%\vspace*{0.5cm}
%\centerline{\it 
%  $\star$ Institut f$\ddot{u}$r Theoretische Physik,
%  ETH Zurich, 8093 Z$\ddot{u}$rich, Switzerland}
\vspace*{0.3cm}
\centerline{\it 
  $\ast$
  Asia Pacific Center for Theoretical Physics, Pohang 37673, Korea }
\vspace*{0.8cm} 
\centerline{\tt ahn@knu.ac.kr, hongjaesung@knu.ac.kr,
  manhea.kim@apctp.org 
  %\qquad  
} 
\vskip2cm

\centerline{\bf Abstract}
\vspace*{0.5cm}

It is known previously that the operator product expansion (OPE)
between the first
${\cal N}=3 $ multiplet and itself contains the second
${\cal N}=3$ multiplet in the supersymmetric coset model.
In this paper, by using their realizations in terms of various fermions,
we compute the four kinds of OPEs between the first and the second
${\cal N}=3$ multiplets for fixed $N$ and $M$ where the group
 of the coset contains $SU(N+M)$.
By supersymmetrizing the above OPEs in ${\cal N}=3$ superspace
and using the various Jacobi identities between the currents,
we determine the ${\cal N}=3$ supersymmetric
OPE between the first and the second
${\cal N}=3$ multiplets completely.
The right hand side of this OPE
contains the various ${\cal N}=3$ multiplets:
the $SO(3)$ singlet ${\cal N}=3$ multiplets
of superspin-$\frac{3}{2},2,3,4$
and  the $SO(3)$ triplet ${\cal N}=3$ multiplets
of superspin-$\frac{5}{2},3,\frac{7}{2}$.
The ${\cal N}=2$ superspace description and the decoupling of
the spin-$\frac{1}{2}$ current of the ${\cal N}=3$ superconformal algebra
are also described.

%\vspace*{4cm}
% \begin{flushright}
%{\it On the occasion of my sixtieth birthday}
%\end{flushright}

\baselineskip=18pt
\newpage
\renewcommand{\theequation}
{\arabic{section}\mbox{.}\arabic{equation}}

\tableofcontents

%%%%%%%%%%%%%%%%%%%%%%%%%%%%%%%%%%%%%%%%%%%%%%%%%%%%%%%%%%%%%%%%%%%%%
%%%%%%%%%%%%%%%%%%%%%%%%%%%%%%%%%%%%%%%%%%%%%%%%%%%%%%%%%%%%%%%%%%%%%%

%%%%%%%%%%%%%%%%%%%%%%%%%%%%%%%%%%%%%%%%%%%%%%%%%%%%%%%%%%%%%%%%%%%%%
%%%%%%%%%%%%%%%%%%%%%%%%%%%%%%%%%%%%%%%%%%%%%%%%%%%%%%%%%%%%%%%%%%%%%%
\section{ Introduction}
%1%%%%%%%%%%%%%%%%%%%%%%%%%%%%%%%%%%%%%%%%%%%%%%%%%%%%%%%%%%%%%%%%%%%%%
%%%%%%%%%%%%%%%%%%%%%%%%%%%%%%%%%%%%%%%%%%%%%%%%%%%%%%%%%%%%%%%%%%%%%

The Virasoro zeromode acting on the primary state,
which corresponds to 
the WZW primary field, is given by the quadratic Casimir operator 
of the finite Lie algebra with some coefficient.
Then the spin of the primary state
is the $\frac{1}{2}$ times 
the quadratic Casimir 
eigenvalues divided by the sum of the level and the
dual Coxeter number of the
finite Lie algebra \cite{BS,DMS}.
The adjoint representation at the level, 
which is equal to the dual Coxeter number,
has the spin-$\frac{1}{2}$.
For the diagonal bosonic coset model \cite{BS}, the
${\cal N}=1$ supersymmetry generator of spin-$\frac{3}{2}$
was initiated by
\cite{Douglas,BBSScon,GS88} and
see also the previous coset constructions in
\cite{GKO,HR,ASS,SS,Ahn1211,Ahn1305,BFK,GHKSS,Ahn1604,Ahn1701,AP1902}
for the various supersymmetric models
\footnote{The other way to
  generalize the bosonic theory to the supersymmetric theory
in two dimensional conformal field theory is to introduce the
complex fermions in the bosonic coset model.}.
In the different type of 
${\cal N}=2$ supersymmetric coset model \cite{KS1,KS2},
its ${\cal N}=3$ version in \cite{CHR1406} has been studied.
Moreover, 
the lowest ${\cal N}=3$ multiplet   
in terms of various fermions has been obtained in \cite{AK1607}.

The coset model in \cite{KS1,KS2} is described as
\bea
\frac{{SU}(N+M)_{k} \times {SO}(2 N M)_1}{{SU}(N)_{k+M} 
\times {SU}(M)_{k+N} 
\times {U}(1)_{NM(N+M)(k+N+M)}}.
\label{KScoset}
\eea
Note that the level $k$ in (\ref{KScoset})
should be equal to the dual Coxeter number
of $SU(N+M)$ for its ${\cal N}=3$ version: $k =N+M$.
The additional third
supersymmetry generator of spin-$\frac{3}{2}$
is obtained from the adjoint fermions of $SU(N+M)$
and the fermions of $SO(2N M)_1$
\cite{CHR1406}.

In this paper,
by using the explicit realizations
for the ${\cal N}=3$ multiplets
in terms of various fermions,
we calculate the four kinds of OPEs between the first and the second
${\cal N}=3$ multiplets for fixed $N$ and $M$ ($N=3, M=2$)
with the help of
the Thielemans package \cite{Thielemans} and a mathematica \cite{mathematica}.
By supersymmetrizing the above OPEs in ${\cal N}=3$ superspace
and using the various Jacobi identities between the currents
as done in \cite{AK1509,AKP1904},
 we determine the ${\cal N}=3$ supersymmetric
OPE between the first and the second
${\cal N}=3$ multiplets completely.
The right hand side of this OPE
contains
the following ${\cal N}=3$ multiplets \footnote{
The  ${\cal N}=3$  superspace 
coordinates  \cite{Ademolloetal,Ademolloetal1,CK,GS,Schoutens88}
are described by
$(Z, \overline{Z}),$ where 
$Z=(z, \theta^i)$ and $\overline{Z} =(\bar{z},$ $\bar{\theta}^i)$,
with $SO(3)$-vector index $i=1,2,3$.
The left covariant spinor derivative 
is 
$ D^i = \theta^i \frac{\pa}{\pa z } +  \frac{\pa}{\pa {\theta^i}}$
satisfying
the anticommutators
$
\{ D^i, D^j \} = 2 \delta^{ij} \frac{\pa}{\pa z}$.
The fermionic coordinate difference for given index $i$ is defined as
 $ \theta_{12}^i = \theta_1^i-\theta_2^i$, and the bosonic
coordinate difference is given by $z_{12} = z_1 -z_2 -
\theta_1^i \theta_2^i$.
We will follow the conventions used in \cite{AK1607}.} 
\bea
{\bar{\bf \Phi }}^{(\frac{3}{2})}(Z)
&= &
\frac{i }{2}\bar{\psi }^{(\frac{3}{2})}(z)+\theta ^{i}\frac{i }{2}
\bar{\phi }^{(2),i}(z)+\theta ^{3-i}\frac{1 }{2}
\bar{\psi }^{(\frac{5}{2}),i}(z)+\theta ^{3-0}\bar{\phi }^{(3)}(z),
\nonumber\\
    {\bar{\bf \Phi }}^{(2)}(Z) &= &
    \frac{i }{2}\bar{\psi }^{(2)}(z)+\theta ^{i}\frac{i }{2}
    \bar{\phi }^{(\frac{5}{2}),i}(z)+\theta ^{3-i}\frac{1 }{2}
    \bar{\psi }^{(3),i}(z)+\theta ^{3-0}\bar{\phi }^{(\frac{7}{2})}(z),
    \nonumber\\
        {\bar{\bf \Phi }}^{(\frac{5}{2}),\alpha }(Z)&= &
        \frac{i }{2}\bar{\psi }^{(\frac{5}{2}),\alpha }(z)+
        \theta ^{i}\frac{i }{2}\bar{\phi }^{(3),i,\alpha }(z)+
        \theta ^{3-i}\frac{1 }{2}\bar{\psi }^{(\frac{7}{2}),i,\alpha }(z)+
        \theta ^{3-0}\bar{\phi }^{(4),\alpha }(z),
        \nonumber\\
{\bar{\bf \Phi }}^{(3),\alpha }(Z)
&=& \frac{i }{2}\bar{\psi }^{(3),\alpha }(z)+\theta ^{i}\frac{i}{2}
\bar{\phi }^{(\frac{7}{2}),i,\alpha }(z)+\theta ^{3-i}\frac{1 }{2}
\bar{\psi }^{(4),i,\alpha }(z)+\theta ^{3-0}
\bar{\phi }^{(\frac{9}{2}),\alpha }(z),
\nonumber\\
{\tilde{\bf \Phi }}^{(3)}(Z)
&=& \frac{i }{2}\tilde{\psi }^{(3)}(z)+\theta ^{i}\frac{i }{2}
\tilde{\phi }^{(\frac{7}{2}),i}(z)+\theta ^{3-i}\frac{1 }{2}
\tilde{\psi }^{(4),i }(z)+\theta ^{3-0}\tilde{\phi }^{(\frac{9}{2}) }(z),
\nonumber\\
{\bar{\bf \Phi }}^{(\frac{7}{2}),\alpha }(Z)
&=& \frac{i }{2}\bar{\psi }^{(\frac{7}{2}),\alpha }(z)+
\theta ^{i}\frac{i }{2}\bar{\phi }^{(4),i,\alpha }(z)+
\theta ^{3-i}\frac{1 }{2}\bar{\psi }^{(\frac{9}{2}),i,\alpha }(z)+
\theta ^{3-0}\bar{\phi }^{(5),\alpha }(z),
\nonumber\\
{\bar{\bf \Phi }}^{(4)}(Z)
&=& \frac{i}{2}\bar{\psi }^{(4)}(z)+\theta ^{i}\frac{i }{2}
\bar{\phi }^{(\frac{9}{2}),i}(z)+\theta ^{3-i}\frac{1 }{2}
\bar{\psi }^{(5),i}(z)+\theta ^{3-0}\bar{\phi }^{(\frac{11}{2})}(z).
\label{list}
\eea
These ${\cal N}=3$ multiplets do not overlap with the ones in
\cite{AK1607}. We do not present the coset field contents for these
currents because we know only for fixed $(N, M)$ values
and they are rather complicated \footnote{The independent terms of the spin-$4$ current
  $\bar{\psi}^{(4)}$ for $(N,M)=(3,2)$ are more than $180000$.}.
%In section $2$,
%
%In section $3$,
%
%In section $4$,

%%%%%%%%%%%%%%%%%%%%%%%%%%%%%%%%%%%%%%%%%%%%%%%%%%%%%%%%%%%%%%%%%%%%%
%%%%%%%%%%%%%%%%%%%%%%%%%%%%%%%%%%%%%%%%%%%%%%%%%%%%%%%%%%%%%%%%%%%%%%
\section{ The OPEs between the first and second ${\cal N}=3$
multiplets in the component approach}
%1%%%%%%%%%%%%%%%%%%%%%%%%%%%%%%%%%%%%%%%%%%%%%%%%%%%%%%%%%%%%%%%%%%%%%
%%%%%%%%%%%%%%%%%%%%%%%%%%%%%%%%%%%%%%%%%%%%%%%%%%%%%%%%%%%%%%%%%%%%%

Let us consider the four types of OPEs
between the currents for fixed $(N,M)=(3,2)$ where the
central charge is $c=9$ \footnote{The central charge is given by $c=\frac{3}{2}
  N M$.}. 
By using the ${\cal N}=3$ supersymmetry, the remaining twelve
types of OPEs can be determined. By using the Jacobi identity
between these OPEs with arbitrary coefficients,
we obtain the final structure constants
in these OPEs in terms of the arbitrary central charge.

%%%%%%%%%%%%%%%%%%%
\subsection{ The OPE between the spin-$\frac{3}{2}$ current and
the spin-$2$ current}
%%%%%%%%%%%%%%%%%%%

In \cite{AK1607}, the lowest components of the
first and second ${\cal N}=3$ multiplet in terms of the coset fields
are known.
Then we can calculate the following OPE
\bea
\psi ^{(\frac{3}{2})}(z)\psi ^{(2)}(w)
& = & \frac{1}{(z-w)^2}\Bigg[-2 i \bar{C}_{\left(\frac{3}{2}\right)
    (2)}^{\left(\frac{3}{2}\right)}
  \bar{\psi }^{\left(\frac{3}{2}\right)}\Bigg](w)
\nonu \\
& + &  \frac{1}{(z-w)}\Bigg[ \frac{1}{3} \pa (\mbox{pole-2})
  %-\frac{2}{3} i
  %\bar{C}_{\left(\frac{3}{2}\right) (2)}^{\left(\frac{3}{2}\right)}
  %\partial \bar{\psi }^{\left(\frac{3}{2}\right)}
  \Bigg](w) + \cdots.
\label{ope1}
\eea
It turns out that there exists a new \footnote{Although the field
  contents of this spin-$\frac{3}{2}$ current are the same as the one
  spin-$\frac{3}{2}$ current appearing in the left hand side of (\ref{ope1}),
  each coefficient appearing in the composite operators is different from
  each other. We have
$
 % \bar{\psi}^{\left( \frac{3}{2} \right)}(z)=\frac{243 }{650\sqrt{2}}
 % ( -\frac{4}{9} J_{1}^{\alpha}+J_{2}^{\alpha}+j^{\alpha} )
 % \Psi^{\alpha}(z)
 % +\frac{3\sqrt{13}}{25}  ( - J_{1}^{\rho}+J_{2}^{\rho}+j^{\rho}
 % ) \Psi^{\alpha}(z)
  6\sqrt{\frac{3}{65}}(\frac{27}{20}
  \sqrt{\frac{3}{130}}(\frac{-4}{9}J_{1}^{\alpha}+
  J_{2}^{\alpha}+j^{\alpha})\Psi^{\alpha}(z)+\frac{13}{10\sqrt{15}}
  (-J_{1}^{\rho}+J_{2}^{\rho}+j^{\rho})\Psi^{\rho}(z))
$ while $
  {\psi}^{\left( \frac{3}{2} \right)}(z)= \frac{3 \sqrt{\frac{3}{13}}}{5}
  ( -\frac{4}{9} J_{1}^{\alpha}+J_{2}^{\alpha}+j^{\alpha} )
  \Psi^{\alpha}(z)
  +\frac{\sqrt{\frac{3}{2}}}{5}  ( - J_{1}^{\rho}+J_{2}^{\rho}+j^{\rho}
  ) \Psi^{\alpha}(z)$ from \cite{AK1607} where the adjoint index $\al$ is
  for $SU(N)$ and the adjoint index $\rho$ is for $SU(M)$.
  The various spin-$1$ currents are defined in Appendix $C$ of \cite{AK1607}.
}
primary current
$ \bar{\psi }^{\left(\frac{3}{2}\right)}$ with the structure
constant $\bar{C}_{\left(\frac{3}{2}\right)
  (2)}^{\left(\frac{3}{2}\right)}$ at the second order pole
of (\ref{ope1}) \footnote{The upper index of the
  structure constant stands for
  the spin of the ${\cal N}=3$ multiplet in (\ref{list}).
  In this paper, there are seven undetermined structure constants
  associated with the ones in (\ref{list}) and they can be fixed
  by calculating the OPEs between each ${\cal N}=3$ multiplet and
itself.}.
At the first order pole its descendant term appears.
It is straightforward to compute all the other components
in the first ${\cal N}=3$ multiplet (\ref{list})
by using Appendix $B$ (the ${\cal N}=3$ primary conditions) of
\cite{AK1607}.

%%%%%%%%%%%%%%%%%%%%
\subsection{ The OPE between the spin-$2$ currents and
the spin-$2$ current}
%%%%%%%%%%%%%%%%%%%%

Now we can move on the next component of the first ${\cal N}=3$ multiplet
and the lowest component of the second ${\cal N}=3$ multiplet
of \cite{AK1607} \footnote{
  We have the ${\cal N}=3$ stress energy tensor as follows:
  ${\bf J}(Z) =
\frac{i }{2} \Psi(z)+\theta ^{i}\frac{i }{2}
J^{i}(z)+\theta ^{3-i}\frac{1 }{2}
G^{i}(z)+\theta ^{3-0} T(z)
  $.}
\bea
\phi ^{(2),i}(z)\psi ^{(2)}(w) & = &
\frac{1}{(z-w)^2} \Bigg[-\frac{2}{3} i
  \bar{C}_{\left(\frac{3}{2}\right)(2)}^{\left(\frac{3}{2}\right)}
  \bar{\phi }^{(2),i}\Bigg](w)
+  \frac{1}{(z-w)}
\Bigg[
  \frac{1}{2} \pa (\mbox{pole-2})\nonu \\
  &+&
  \bar{C}_{\left(\frac{3}{2}\right)(2)}^{\left(\frac{3}{2}\right)}
  \left( \frac{ 27 i   (c+1) }{c (c+3)(c-3)}
  \Psi J^i \bar{\psi }^{(\frac{3}{2})} -\frac{ 9 i
    (c+1) }{(c+3)(c-3)}G^i \bar{\psi }^{(\frac{3}{2})}
  \right.  \nonumber\\
  &- & \left. \frac{   (c-15)  }{(c+3)(c-3)}
  \varepsilon^{ijk} J^j\bar{\phi }^{(2),k}+
  \frac{i   (5c-3)  }{ (c+3)(c-3)} \partial
  \bar{\phi }^{(2),i} \right) \nonumber \\
  &+& \bar{C}_{(3)(2)}^{(2)} \left(
  \frac{ c   }{10(c-3)}\bar{\psi }^{(3),i}-
  \frac{3  }{10 (c-3)}\Psi \bar{\phi   }^{(\frac{5}{2}),i}
  -\frac{6  }{5 (c-3)}J^i \bar{\psi }^{(2)} \right)
  \nonumber\\
  & -& \bar{C}_{\left(\frac{5}{2}\right)(2)}^{\left( \frac{5}{2}\right)}
  \left( \frac{  c }{3(c-1)}\delta ^{i\alpha }
  (T^j)^{\alpha \beta } \bar{\phi }^{(3),j,\beta}+
  \frac{2  }{(c-1)}\Psi \bar{\psi   }^{(\frac{5}{2}),\alpha=i}
  \right)  \nonumber \\
  &- &  2 i \bar{C}_{(2)(2)}^{(3)} \bar{\psi }^{(3),\alpha=i}
  \Bigg](w) + \cdots.
\label{ope2}
\eea
After subtracting the descendant term in the first order pole of
(\ref{ope2}), we are left
with three primary operators
and a new primary current of spin-$3$.
First of all, it is not clear how we can consider the last two
primary operators, $\bar{C}_{(3)(2)}^{(2)}$ and
$\bar{C}_{\left(\frac{5}{2}\right)(2)}^{\left( \frac{5}{2}\right)}$ terms,
at the first order pole.
The way we do here is to calculate the highest order poles appearing
in next two subsections, identify the two new lowest components
of the second and third ${\cal N}=3$ multiplet and obtain
all the other components in each ${\cal N}=3$ multiplet explicitly.
After that, we return to the first order pole of (\ref{ope2}) by
allowing us to have 
all the possible composite operator terms of spin-$3$ and then it turns out
that
we are left with the first order pole above.

Note that there appears the dependence on the $SO(3)$ generators
\footnote{Explicitly, we have the nonzero components of these 
  generators as follows \cite{AK1607}: $(T^1)^{23}= -(T^1)^{32}=-i$,
  $(T^2)^{13}= -(T^2)^{31}=i$ and  $(T^3)^{12}= -(T^3)^{21}=-i$
  satisfying the algebra $[T^i, T^j] = i \, \epsilon^{i j k} \, T^k$.}
$T^i$ where $i=1,2,3$ compared with the work of \cite{AK1607}.
The two indices $j$ and $\beta$ coming 
from $(T^j)^{\alpha \beta}$ and $\bar{\phi }^{(3),j,\beta}$
are contracted with each other and the index $\alpha$ becomes
the $SO(3)$ vector representation due to the free index $i$ in both sides.
Although we have considered all the possible terms by including
the nonlinear terms in the ${\cal N}=3$ multiplets,
the above results imply that the nonlinear terms come from the
components of ${\cal N}=3$ stress energy tensor 
and the components
of ${\cal N}=3$ multiplets.
In other words, there are no nonlinear terms in the 
components of ${\cal N}=3$ multiplets.

Therefore, we have obtained the new lowest spin-$3$ component of
the fourth ${\cal N}=3$ multiplet in (\ref{list})
and the remaining components can be determined by using the
${\cal N}=3$ supersymmetry currents of the ${\cal N}=3$
stress tensor as before.

%%%%%%%%%%%%%%%%
\subsection{ The OPE between the spin-$\frac{5}{2}$ currents and
the spin-$2$ current}
%%%%%%%%%%%%%%%%%

From the third component of the first ${\cal N}=3$ multiplet
and the lowest component of the second ${\cal N}=3$
multiplet,
we obtain the following OPE
\bea
&& \psi^{(\frac{5}{2}),i}(z)\psi^{2}(w)=
\frac{1}{(z-w)^2}
\Bigg[-\bar{C}_{\left(\frac{3}{2}\right)(2)}^{\left(\frac{3}{2}\right)}
  \left( \frac{i (c-15)  }{3 (c-3)}\bar{\psi }^{\left(\frac{5}{2}\right),i}
  +\frac{i    (c+9) }{c(c-3) }\Psi \bar{\phi }^{(2),i} \right. 
  \nonu \\
  &&  +  \left. 
  \frac{ 3 i  (5 c-3) }{c(c-3) }J^i \bar{\psi }^{\left(\frac{3}{2}\right)}
  \right) +   \frac{1 }{2}\bar{C}_{(3)(2)}^{(2)}
  \bar{\phi }^{\left(\frac{5}{2}\right),i}+2
  \bar{C}_{\left(\frac{5}{2}\right)(2)}^{\left(\frac{5}{2}\right)}
  \bar{\psi   }^{\left(\frac{5}{2}\right),\alpha=i} \Bigg](w)
\nonumber \\
 && +  \frac{1}{(z-w)}\Bigg[  \frac{3}{5}
  \partial ( \mbox{pole-2} ) 
  + \bar{C}_{\left(\frac{3}{2}\right)(2)}^{\left(\frac{3}{2}\right)}
  \Bigg( -\frac{27 i  (c+1) }{c (c+3)(c-3)}\Psi G^i
  \bar{\psi }^{\left(\frac{3}{2}\right)}  \nonumber \\
  && - \frac{1 }{(5 c+6) (c+3)(c-3)}\varepsilon^{ijk}
  ( \vphantom{\frac{2 i  \left(19c^2-18 c-45\right)   }{5}}
  18    (c+1)(c-1) G^j \bar{\phi }^{(2),k} 
   -   \frac{2 i  \left(19c^2-18 c-45\right)   }{5}
  \varepsilon^{jkl} \partial \bar{\psi }^{\left(\frac{5}{2}\right),l}
  ) \nonumber \\
  & & - \frac{2  \left(c^2-18 c-27\right) }{(5 c+6) (c+3)(c-3)}
  \varepsilon^{ijk} J^j \bar{\psi   }^{\left(\frac{5}{2}\right),k}
  \nonumber \\
  & & - \frac{i}{(c+3)(c-3)(5c+6)}( \vphantom{\frac{1}{2}}
  c(5c-27) \partial \Psi  \bar{\phi }^{(2),i} 
  -  \frac{(5c^3+153c^2+396c+216)}{5c} \partial
  (\Psi  \bar{\phi }^{(2),i}) ) \nonumber \\
  &&  -  \frac{12 i (5c^3-18c-9)}{c(c+3)(c-3)(5c+6)}
  (  \partial J^i \bar{\psi }^{\left(\frac{3}{2}\right)}-
  \frac{2}{5} \partial (J^i \bar{\psi }^{\left(\frac{3}{2}\right)})
  ) +   \frac{9   (c+1) }{c (c+3)(c-3)}
  \varepsilon^{ijk} \Psi J^j \bar{\phi }^{(2),k}  \Bigg)
  \nonumber\\
  & & + \bar{C}_{(3)(2)}^{(2)} \Bigg( -\frac{18 }{
    5 (c-3) (2 c+9)}\Psi J^i \bar{\psi }^{(2)} +\frac{3  }{
    10 (c-3)}\Psi \bar{\psi   }^{(3),i} 
  \nonumber \\
  & & - \frac{1 }{5 (c-3) (2 c+9)} ( 6  (c+9) G^i
  \bar{\psi }^{(2)} - 18 \partial \bar{\phi }^{\left(\frac{5}{2}\right),i}
  ) -  \frac{3 i (2 c+3) }{5 (c-3) (2 c+9)}
  \varepsilon^{ijk} J^j \bar{\phi   }^{\left(\frac{5}{2}\right),k}
  \Bigg) \nonu \\
&&
  - \bar{C}_{\left(\frac{5}{2}\right)(2)}^{\left(\frac{5}{2}\right)}
  \Bigg( \frac{   1}{(c-1)}\delta ^{i\alpha }(T^j)^{\alpha \beta }
  \Psi \bar{\phi }^{(3),j,\beta}
  +   \frac{  1 }{7}\delta ^{i\alpha }(T^j)^{\alpha \beta }
  ( \bar{\psi }^{\left(\frac{7}{2}\right),j,\beta} +
  \frac{1}{5} (T^j)^{ \beta \gamma} \partial
  \bar{\psi   }^{\left(\frac{5}{2}\right),\gamma} ) \Bigg)
  \nonumber \\
  && + \bar{C}_{(2)(2)}^{(3)} \Bigg( \frac{6 i  (c-1)  }{
    (7 c-6)}\delta ^{i\alpha }(T^j)^{\alpha \beta }
  \bar{\phi }^{  \left(\frac{7}{2}\right),j,\beta} -
  \frac{6 i }{(7 c-6)} \Psi \bar{\psi }^{(3),\alpha=i} \Bigg)
  \nonumber \\
  && +  \frac{1   }{3}\tilde{C}_{(3)(2)}^{(3)}
  \tilde{\phi }^{\left(\frac{7}{2}\right),i} +2
  \bar{C}_{\left(\frac{5}{2}\right)(2)}^{\left(\frac{7}{2}\right)}
  \bar{\psi }^{\left(\frac{7}{2}\right),\al=i} \Bigg](w)+ \cdots.
\label{ope3}
\eea
As decribed in previous section, we
have found the second order pole of the above OPE (\ref{ope3})
by calculating the highest order pole of the next OPE in next subsection
and identifying the second ${\cal N}=3$ multiplet.
In other words, the second order pole consists of
a primary operator, a primary spin-$\frac{5}{2}$ current
which belongs to the second ${\cal N}=3$ multiplet
and other kind of spin-$\frac{5}{2}$ currents
which is the component of the third ${\cal N}=3$ multiplet in
(\ref{list}).

At the first order pole, there are four kinds of primary operators,
a primary spin-$\frac{7}{2}$ current belonging to the
fifth ${\cal N}=3$ multiplet in (\ref{list})
and other primary spin-$\frac{7}{2}$
currents which are the lowest components of the sixth ${\cal N}=3$
multiplet in (\ref{list}).
Also note that in the term of $\bar{C}_{(2)(2)}^{(3)}$  in (\ref{ope3}),
we observe the presence of the fourth ${\cal N}=3$ multiplet.

%%%%%%%%%%%%%%
\subsection{ The OPE between the spin-$3$ current and
the spin-$2$ current}
%%%%%%%%%%%%%%

The final most complicated OPE
can be summarized by
\bea
&& \phi^{(3)}(z)\psi^{(2)}(w)  = 
\frac{1}{(z-w)^3}
\Bigg[-  \frac{6 i }{c}
  \bar{C}_{\left(\frac{3}{2}\right)(2)}^{\left(\frac{3}{2}\right)}
  \Psi \bar{\psi   }^{\left(\frac{3}{2}\right)} +
  \bar{C}_{(3)(2)}^{(2)} \bar{\psi }^{(2)} \Bigg](w)
\quad \quad \quad \quad \quad \quad \quad \quad \quad \nonumber \\
&& +  \frac{1}{(z-w)^2} \Bigg[ \frac{3   }{4} \partial
  (\mbox{pole-3}) -
  \bar{C}_{\left(\frac{3}{2}\right)(2)}^{\left(\frac{3}{2}\right)}
  \left(\frac{2 i  (7 c+15) }{(c-3) (5 c+6)}J^i\bar{\phi }^{(2),i}
  \right.  \nonumber \\
  & & +  \left. \frac{8 i  (8 c+3)}{(c-3) (5   c+6)}
  (  \partial \Psi \bar{\psi }^{\left(\frac{3}{2}\right)}-
  \frac{1}{4}  \partial  (
  \Psi \bar{\psi }^{\left(\frac{3}{2}\right)}   )  )
  +\frac{i \left(c^2-25 c-42\right)  }{(c-3) (5   c+6)}
  \bar{\phi }^{(3)} \right) \nonumber \\
  & & +  \frac{3 }{7}
  \bar{C}_{\left(\frac{5}{2}\right)(2)}^{\left(\frac{5}{2}\right)}
  \bar{\phi }^{(3),i,\alpha=i} +\tilde{C}_{(3)(2)}^{(3)}
  \tilde{\psi }^{(3)} \Bigg](w)
%%%%%%%%%%%%%%%%%%%%%%%%%%%%%%%%%%%%%%%%%%%%%%%%%%%%%%%%%
+  \frac{1}{(z-w)} \Bigg[ \frac{3}{10} \partial^2
  ( \mbox{pole-3} ) +\frac{2}{3} \partial
  ( \mbox{pole-2} - \frac{3}{4} \partial (\mbox{pole-3})
  )  \nonumber \\
  & & +  \bar{C}_{\left(\frac{3}{2}\right)(2)}^{\left(\frac{3}{2}\right)}
  \left( \frac{3 i    (c-12) (c+1) }{ c(c-3)  (c+3)^2}
  ( \Psi J^i \bar{\psi }^{\left(\frac{5}{2}\right),i}-
  \Psi G^i \bar{\phi }^{\left(2\right),i} ) \right.
  \nonumber \\
  & & +  \frac{135 i  (c+1) }{c (c-3)  (c+3)^2}
  \Psi J^i J^i \bar{\psi }^{\left(\frac{3}{2}\right)}-
  \frac{i  (c-12) (c+1) }{(c-3) (c+3)^2}
  ( G^i \bar{\psi }^{\left(\frac{5}{2}\right),i} -
  2 \partial \bar{\phi}^{(3)} ) \nonumber \\ 
  & & -  \frac{1 }{(c-3) (c+3)^2} ( 45 i (c+1)
  J^i G^i \bar{\psi }^{\left(\frac{3}{2}\right)} -
  \frac{2 i c (c+33) }{3}J^i  \partial
  \bar{\phi }^{(2),i} ) \nonumber \\
  & & -  \frac{i \left(4 c^2-3 c-135\right) }{3(c-3) (c+3)^2}
  \partial J^i \bar{\phi   }^{(2),i} -
  \frac{i (16c^3+21c^2+153c-108)}{2c(c+3)^2(c-3)}
  \partial^2 \Psi \bar{\psi }^{\left(\frac{3}{2}\right)}
  \nonumber \\
  & & \left. +  \frac{4i(4c^2+3c+63)}{3(c+3)^2(c-3)}\partial
  ( \partial \Psi  \bar{\psi }^{\left(\frac{3}{2}\right)})
  -   \frac{i \left(8 c^3-21 c^2+423 c+324\right) }{
    15 c (c-3)(c+3)^2}  \partial^2 ( \Psi
  \bar{\psi }^{\left(\frac{3}{2}\right)} )
  \right) \nonumber \\
  & & +  \bar{C}_{(3)(2)}^{(2)} \left( -\frac{18
    \left(8 c^2+53 c+57\right) }{5 (c-3) (2 c+9)
    \left(c^2+26 c+9\right)}J^i J^i \bar{\psi }^{(2)}
  \right. \nonumber \\
  & & +  \frac{3  \left(2 c^3+7 c^2+63 c+90\right)  }{
    5 (c-3) (2 c+9) \left(c^2+26 c+9\right)}J^i \bar{\psi }^{(3),i}
  +  \frac{12  c \left(4 c^2+9 c-99\right) }{
    5 (c-3) (2 c+9) \left(c^2+26   c+9\right)}
  T\bar{\psi }^{(2)}\nonumber \\
  & & -  \frac{9 \left(6 c^3+17 c^2-105 c-36\right)}{10 (c-3)
    (2 c+9) \left(c^2+26 c+9\right)}  \partial^2
  \bar{\psi }^{(2)}   -  \frac{27
    \left(6 c^2+21 c-1\right) }{10 (c-3)
    (2 c+9) \left(c^2+26 c+9\right)}\Psi J^i
  \bar{\phi }^{\left(\frac{5}{2}\right),i} \nonumber \\
  & & \left. +  \frac{3  \left(14 c^3+49 c^2-129 c-180\right) }{
    10   (c-3) (2 c+9) \left(c^2+26 c+9\right)}G^i
  \bar{\phi }^{\left(\frac{5}{2}\right),i} -
   \frac{18  \left(4 c^2+9 c-99\right) }{
    5 (c-3) (2 c+9) \left(c^2+26 c+9\right)}
  \partial \Psi \Psi \bar{\psi   }^{(2)} \right)
  \nonumber \\
  & & + \bar{C}_{\left(\frac{5}{2}\right)(2)}^{\left(\frac{5}{2}\right)}
  \left(-\frac{2 c    }{(c-1) (c+15)} \delta ^{i\alpha }
  (T^j)^{\alpha \beta } J^i \bar{\phi }^{(3),j,\beta}\right.
  -  \frac{30    }{(c-1) (c+15)}\Psi J^i
  \bar{\psi }^{\left(\frac{5}{2}\right),\alpha=i} \nonumber \\
  & & +  \left. \frac{6c   }{(c-1)   (c+15)} ( G^i
  \bar{\psi }^{\left(\frac{5}{2}\right),\alpha=i}-\frac{1 }{3}
  \partial \bar{\phi   }^{(3),i,\alpha=i} )
  \right) \nonumber \\ 
  & & -  \bar{C}_{(2)(2)}^{(3)} \left( \frac{3 i    c(c-2) }{
    2 (c+3) (7 c-6)}\bar{\psi }^{(4),i,\alpha=i}  +
  \frac{9 i  (c-2) }{2 (c+3) (7 c-6)}\Psi
  \bar{\phi }^{\left(\frac{7}{2}\right),i,\alpha=i} \right.
  \nonumber \\
  & & +   \left. \frac{12 i  (4 c-3) }{
    (c+3) (7 c-6)}J^i \bar{\psi }^{(3),\alpha=i}  \right)
  +\frac{4 }{9}
  \bar{C}_{\left(\frac{5}{2}\right)(2)}^{\left(\frac{7}{2}\right)}
  \bar{\phi }^{(4),i,\alpha=i} +\bar{C}_{(3)(2)}^{(4)}
  \bar{\psi }^{(4)} \Bigg](w) + \cdots.
\label{ope4}
\eea
As explained before, the third  order pole of (\ref{ope4})
contains the lowest component of the second ${\cal N}=3$
multiplet in (\ref{list}).
The component of fifth ${\cal N}=3$ multiplet
appears at the second order pole.
We arrive at the lowest components of the
sixth and seventh ${\cal N}=3$ multiplets
in the first order pole of (\ref{ope4}).
The two quasiprimary operators with
$\bar{C}_{\left(\frac{3}{2}\right)(2)}^{\left(\frac{3}{2}\right)}$
and $\bar{C}_{(3)(2)}^{(2)}$
at the first order pole appear while the next two operators
are primary.

Therefore, we
have obtained the new seven ${\cal N}=3$ multiplets
in (\ref{list}) by checking the presence of
lowest components of those multiplets.

%%%%%%%%%%%%%%
\subsection{ The
  ${\cal N}=3$ superspace and the Jacobi identity}
%%%%%%%%%%%%%%

Because the neccessary four kinds of OPEs in the previous subsections
are found explicitly, we can use ${\cal N}=3$ supersymmetry
in the coset model we are describing to determine the
remaining OPEs between the first and the second ${\cal N}=3$ multiplets.
That is, we simply generalize
the above four kinds of OPEs to the complete OPEs
in ${\cal N}=3$ superspace by the following replacement 
\bea
{\psi}^{\left(\Delta \right), \alpha} & \rightarrow &
-2 i {\bf \Phi}^{(\Delta),\alpha},
\nonu \\
{\phi}^{\left(\Delta + \frac{1}{2} \right),i, \alpha} & \rightarrow &
-2 i D^{i} {\bf \Phi}^{(\Delta),\alpha}, \nonu \\
{\psi}^{\left(\Delta +1 \right), i,\alpha} & \rightarrow &
-2 D^{3-i} {\bf \Phi}^{(\Delta),\alpha}, \nonu \\
{\phi}^{\left(\Delta + \frac{3}{2} \right), \alpha} & \rightarrow &
-D^{3-0} {\bf \Phi}^{(\Delta),\alpha},
\label{comptofull}
\eea
together with the appropriate fermionic coordinates and
the distance between the point $Z_1$ and the point $Z_2$
\footnote{Similarly, we have, for the ${\cal N}=3$ multiplets
  in (\ref{list}),
$\bar{\psi}^{\left(\Delta \right), \alpha}  \rightarrow 
-2 i \bar{\bf \Phi}^{(\Delta),\alpha}$,
$
\bar{\phi}^{\left(\Delta + \frac{1}{2} \right),i, \alpha}  \rightarrow 
-2 i D^{i} \bar{\bf \Phi}^{(\Delta),\alpha}$,
$\bar{\psi}^{\left(\Delta +1 \right), i,\alpha}  \rightarrow 
-2 D^{3-i} \bar{\bf \Phi}^{(\Delta),\alpha}$ and 
$\bar{\phi}^{\left(\Delta + \frac{3}{2} \right), \alpha}  \rightarrow 
-D^{3-0} \bar{\bf \Phi}^{(\Delta),\alpha}$.}.

Then we can read off the remaining OPEs by taking the ${\cal N}=3$
superderivatives and putting the fermionic coordinates to zero.
Because the complete OPEs are known in terms of the ${\cal N}=3$
stress energy tensor and the various ${\cal N}=3$ multiplets explicitly,
we can put the arbitrary coefficients in the right hand sides of
the complete OPEs and the number of unknown coefficients is given by
sixty seven. By using the Jacobi identity between the currents
of the ${\cal N}=3$ stress energy tensor and the currents of the
${\cal N}=3$ multiplets, all the structure constants are fixed
in terms of the central charge. Some of them appear in the previous
OPEs in (\ref{ope1}), (\ref{ope2}), (\ref{ope3}) and (\ref{ope4}).

Then by writing down these OPEs in ${\cal N}=3$ superspace,
we have the final form in next section. In Appenidx $A$,
we present the ${\cal N}=3$ multiplets in (\ref{list}) by using its
${\cal N}=2$ superspace approach. In Appendix $B$,
by decoupling the spin-$\frac{1}{2}$ current of the
${\cal N}=3$ superconformal algebra, the corresponding components
of ${\cal N}=3$ multiplets are described. 

%%%%%%%%%%%%%%%%%%%%%%%%%%%%%%%%%%%%%%%%%%%%%%%%%%%%%%%%%%%%%%%%%%%%%
%%%%%%%%%%%%%%%%%%%%%%%%%%%%%%%%%%%%%%%%%%%%%%%%%%%%%%%%%%%%%%%%%%%%%%
\section{ The ${\cal N}=3$ supersymmetric OPE}
%1%%%%%%%%%%%%%%%%%%%%%%%%%%%%%%%%%%%%%%%%%%%%%%%%%%%%%%%%%%%%%%%%%%%%%
%%%%%%%%%%%%%%%%%%%%%%%%%%%%%%%%%%%%%%%%%%%%%%%%%%%%%%%%%%%%%%%%%%%%%

Therefore, by reexpressing all the component OPEs found in previous
section in ${\cal N}=3$ supersymmetric way,
we arrive at the final ${\cal N}=3$ supersymmetric
OPE can be described as
\bea
&&    {\bf \Phi}^{\left(\frac{3}{2}\right)}(Z_1)
    {\bf \Phi}^{(2)}(Z_2) =
    \frac{1}{z_{12}^{2}} \Bigg[
      \bar{C}_{(\frac{3}{2})(2)}^{(\frac{3}{2})}
      {\bar{\bf \Phi }}^{(\frac{3}{2})} \Bigg](Z_2)+
    \frac{\theta_{12}^{3-0}}{z_{12}^{3}} \Bigg[ -\frac{12}{c}
      \bar{C}_{(\frac{3}{2})(2)}^{(\frac{3}{2})} {\bf J}
          {\bar{\bf \Phi }}^{(\frac{3}{2})} +
          \bar{C}_{(3)(2)}^{(2)} {\bar{\bf \Phi }}^{(2)}
          \Bigg] (Z_2)\nonumber\\
   && +  \frac{\theta_{12}^{i}}{z_{12}^{2}} \Bigg[\frac{1}{3}
      \bar{C}_{(\frac{3}{2})(2)}^{(\frac{3}{2})}D^{i}
          {\bar{\bf \Phi }}^{(\frac{3}{2})} \Bigg](Z_2)
 +  \frac{\theta_{12}^{3-i}}{z_{12}^{2}} \Bigg[
      \bar{C}_{(\frac{3}{2})(2)}^{(\frac{3}{2})}
      \left( -\frac{(c-15)}{6(c-3)}D^{3-i}
      {\bar{\bf \Phi }}^{(\frac{3}{2})} -\frac{(c+9)}{c(c-3)}
      {\bf J}D^{i}{\bar{\bf \Phi }}^{(\frac{3}{2})} \right. \nonu \\
      && -   \left.
      \frac{3(5c-3)}{c(c-3)}D^{i}
      {\bf J} {\bar{\bf \Phi }}^{(\frac{3}{2})} \right)
       +  \frac{1}{4}\bar{C}_{(3)(2)}^{(2)} D^{i}
      {\bar{\bf \Phi }}^{(2)}+
      \bar{C}_{(\frac{5}{2})(2)}^{(\frac{5}{2})}\delta ^{i\alpha }
      {\bar{\bf \Phi }}^{(\frac{5}{2}),\alpha}\Bigg](Z_2)
    \nonumber\\
    && +  \frac{\theta_{12}^{3-0}}{z_{12}^{2}} \Bigg[ \frac{3}{4}
      \partial ( \frac{\theta ^{3-0}}{z_{12}^{3}}-\mbox{term})
      -\bar{C}_{(\frac{3}{2})(2)}^{(\frac{3}{2})}
      \left( \frac{4(7c+15)}{(c-3)(5c+6)}
      D^{i} {\bf J} D^{i} {\bar{\bf \Phi }}^{(\frac{3}{2})}
      \right.  \nonumber\\
      && \left. +  \frac{16(8c+3)}{(c-3)(5c+6)}
      (  \partial {\bf J} {\bar{\bf \Phi }}^{(\frac{3}{2})}-
      \frac{1}{4} \partial ({\bf J} {\bar{\bf \Phi }}^{(\frac{3}{2})})
      )   +  \frac{(c^2-25c-42)}{2(c-3)(5c+6)}
      D^{3-0} {\bar{\bf \Phi }}^{(\frac{3}{2})} \right)
      \nonu \\
      && +  \frac{3}{7}\bar{C}_{(\frac{5}{2})(2)}^{(\frac{5}{2})}
      \delta ^{i\alpha }D^{i} {\bar{\bf \Phi }}^{(\frac{5}{2}),\alpha}+
      \tilde{C}_{(3)(2)}^{(3)} {\tilde{\bf \Phi }}^{(3)}
      \Bigg] (Z_{2}) 
    %%%%%%%%%%%%%%%%%%%%%%%%%%%%%%%%%%
       +  \frac{1}{z_{12}} \Bigg[
      \frac{1}{3}
      \partial (\frac{1}{z_{12}^{2}}-\mbox{term}) \Bigg](Z_2)\nonu \\
       && +  \frac{\theta_{12}^{i}}{z_{12}} \Bigg[
         \frac{1}{2}
      \partial (\frac{\theta_{12}^i}{z_{12}^{2}}-\mbox{term}) +
      \bar{C}_{(\frac{3}{2})(2)}^{(\frac{3}{2})}\left(
      \frac{54(c+1)}{c(c+3)(c-3)}
      {\bf J}D^{i}{\bf J}{\bar{\bf \Phi }}^{(\frac{3}{2})}-\frac{9(c+1)}{
        (c+3)(c-3)}D^{3-i}{\bf J}{\bar{\bf \Phi }}^{(\frac{3}{2})}
      \right.  \nonumber\\
      && -  \left. \frac{(c-15)}{(c+3)(c-3)}\varepsilon ^{ijk}D^{j}
      {\bf J}D^{k}{\bar{\bf \Phi }}^{(\frac{3}{2})} -\frac{
        (5c-3)}{2(c+3)(c-3)} \partial{ D^{i}
        {\bar{\bf \Phi }}^{(\frac{3}{2})}} \right) \nonumber \\
      &&- \bar{C}_{(3)(2)}^{(2)}\left( -\frac{c}{20(c-3)}D^{3-i}
      {\bar{\bf \Phi }}^{(2)}+\frac{3}{10(c-3)}{\bf J}D^{i}
      {\bar{\bf \Phi }}^{(2)}+\frac{6}{5(c-3)}D^{i}{\bf J}
      {\bar{\bf \Phi }}^{(2)} \right) \nonumber \\
      && - \bar{C}_{(\frac{5}{2})(2)}^{(\frac{5}{2})}\left( \frac{i c}{6(c-1)}
      \delta ^{i\alpha }(T^j)^{\alpha \beta }D^{j}
      {\bar{\bf \Phi }}^{(\frac{5}{2}),\beta } +\frac{2}{(c-1)}
      \delta ^{i\alpha }{\bf J}{\bar{\bf \Phi }}^{(\frac{5}{2}),\alpha}
      \right)   +   \bar{C}_{(2)(2)}^{(3)}\delta ^{i\alpha }
            {\bar{\bf \Phi }}^{(3),\alpha} \Bigg](Z_2)\nonu \\
     && +  \frac{\theta_{12}^{3-i}}{z_{12}} \Bigg[\frac{3}{5}  \partial
      (\frac{\theta ^{3-i}}{z_{12}^{2}}-\mbox{term} )+
      \bar{C}_{(\frac{3}{2})(2)}^{(\frac{3}{2})}
      \left( \frac{54(c+1)}{c(c+3)(c-3)}{\bf J}D^{3-i}{\bf J}
      {\bar{\bf \Phi }}^{(\frac{3}{2})} \right.  \nonumber \\
      && + \frac{\varepsilon ^{ijk}}{(c+3)(c-3)(5c+6)} (
      \vphantom{\frac{1}{2}}18(c+1)(c-1)D^{3-j}{\bf J}D^{k}
      {\bar{\bf \Phi }}^{(\frac{3}{2})}  \nonumber\\
      && +   \frac{(19c^2-18c-45)}{5}\varepsilon ^{jkl} \partial
      D^{3-l}{\bar{\bf \Phi }}^{(\frac{3}{2})} )
      + \frac{2(c^2-18c-27)}{(c+3)(c-3)(5c+6)}\varepsilon ^{ijk}D^{j}
      {\bf J}D^{3-k}{\bar{\bf \Phi }}^{(\frac{3}{2})} \nonumber \\
      && - \frac{1}{(c+3)(c-3)(5c+6)}( \vphantom{\frac{1}{2}}
      c(5c-27) \partial {\bf J} D^{i}{\bar{\bf \Phi }}^{(\frac{3}{2})}
       -   \frac{(5c^3+153c^2+396c+216)}{5c} \partial
      ({\bf J}D^{i}{\bar{\bf \Phi }}^{(\frac{3}{2})}) ) \nonumber \\
      && \left. - \frac{12(5c^3-18c-9)}{c(c+3)(c-3)(5c+6)}(
      \partial D^{i}{\bf J} {\bar{\bf \Phi }}^{(\frac{3}{2})}-
      \frac{2}{5} \partial (D^{i}{\bf J}{\bar{\bf \Phi }}^{(\frac{3}{2})})
      ) -   \frac{18(c+1)}{c(c+3)(c-3)}\varepsilon ^{ijk}
      {\bf J}D^{j}{\bf J}D^{k}{\bar{\bf \Phi }}^{(\frac{3}{2})}
      \right)\nonumber \\
      && + \bar{C}_{(3)(2)}^{(2)} \left( \frac{36}{5(c-3)(2c+9)}
      {\bf J}D^{i}{\bf J}{\bar{\bf \Phi }}^{(2)} -\frac{3}{10(c-3)}
      {\bf J}D^{3-i}{\bar{\bf \Phi }}^{(2)} \right. \nonumber \\
      && \left. -  \frac{1}{5(2c+9)(c-3)} ( -6(c+9)D^{3-i}
      {\bf J}{\bar{\bf \Phi }}^{(2)}-9 \partial D^{i}
      {\bar{\bf \Phi }}^{(2)} ) -  \frac{3(2c+3)}{5(2c+9)(c-3)}\varepsilon ^{ijk}D^{j}
      {\bf J}D^{k}{\bar{\bf \Phi }}^{(2)} \right) \nonumber\\
      && \left.
      + \bar{C}_{(\frac{5}{2})(2)}^{(\frac{5}{2})} \left( \frac{i }{(c-1)}
      \delta ^{i\alpha }(T^j)^{\alpha \beta }
      {\bf J}D^{j}{\bar{\bf \Phi }}^{(\frac{5}{2}),\beta } \right.
      -   \frac{i }{14}\delta ^{i\alpha }(T^j)^{\alpha \beta }
      ( -D^{3-j}{\bar{\bf \Phi }}^{(\frac{5}{2}),\beta }-
      \frac{i }{5}(T^j)^{\beta \gamma } \partial
      {\bar{\bf \Phi }}^{(\frac{5}{2}),\gamma  } ) \right) \nonumber\\
      && + \bar{C}_{(2)(2)}^{(3)} \left( \frac{3 i (c-1)}{(7c-6)}
      \delta ^{i\alpha }(T^j)^{\alpha \beta }D^{j}
      {\bar{\bf \Phi }}^{(3),\beta }-\frac{6}{(7c-6)}\delta ^{i\alpha }
      {\bf J}{\bar{\bf \Phi }}^{(3),\alpha} \right) \nonumber\\
      && + \frac{1}{6}\tilde{C}_{(3)(2)}^{(3)}D^{i}{\tilde{\bf \Phi }}^{(3)}+
      \bar{C}_{(\frac{5}{2})(2)}^{(\frac{7}{2})}\delta ^{i\alpha }
      {\bar{\bf \Phi }}^{(\frac{7}{2}),\alpha}    \Bigg](Z_2)
    \nonumber\\
  &&  +  \frac{\theta_{12}^{3-0}}{z_{12}} \Bigg[\frac{3}{10}
     \partial^2 (\frac{\theta ^{3-0}}{z_{12}^{3}}-\mbox{term}) +
     \frac{2}{3}
     \partial \left(
(\frac{\theta ^{3-0}}{z_{12}^{2}}-\mbox{term})
     -\frac{3}{4} \partial
     ( \frac{\theta ^{3-0}}{z_{12}^{3}}-\mbox{term} ) \right)
     \nonumber \\
     & &+ \bar{C}_{(\frac{3}{2})(2)}^{(\frac{3}{2})} \left(
      -\frac{12(c+1)(c-12)}{c(c-3)(c+3)^2} (
      {\bf J}D^{i}  {\bf J}D^{3-i} {\bar{\bf \Phi }}^{(\frac{3}{2})}-
      {\bf J} D^{3-i} {\bf J} D^{i} {\bar{\bf \Phi }}^{(\frac{3}{2})}
      )  \right. \nonumber \\
      && -  \frac{1080(c+1)}{c(c-3)(c+3)^2} {\bf J}D^{i} {\bf J} D^{i}
      {\bf J}
      {\bar{\bf \Phi }}^{(\frac{3}{2})} +
      \frac{2(c+1)(c-12)}{(c-3)(c+3)^2} ( D^{3-i}
      {\bf J} D^{3-i} {\bar{\bf \Phi }}^{(\frac{3}{2})}+
      \frac{1}{2} \partial D^{3-0} {\bar{\bf \Phi }}^{(\frac{3}{2})} )
      \nonumber\\
      && - \frac{1}{(c-3)(c+3)^2} (-180(c+1)D^{i}{\bf J} D^{3-i}
      {\bf J}
      {\bar{\bf \Phi }}^{(\frac{3}{2})} -
      \frac{4c(c+33)}{3}D^{i} {\bf J} \partial
      D^{i} {\bar{\bf \Phi }}^{(\frac{3}{2})} ) \nonu \\
      && -
      \frac{2(4c^2-3c-135)}{3(c+3)^2(c-3)} \partial D^{i}
           {\bf J} D^{i}{\bar{\bf \Phi }}^{(\frac{3}{2})}-
           \frac{(16c^3+21c^2+153c-108)}{c(c+3)^2(c-3)} \partial^2
      {\bf J} {\bar{\bf \Phi }}^{(\frac{3}{2})}\nonumber\\
      && +  \left. \frac{8(4c^2+3c+63)}{3(c+3)^2(c-3)} \partial
      ( \partial {\bf J} {\bar{\bf \Phi }}^{(\frac{3}{2})})-
      \frac{2(8c^3-21c^2+423c+324)}{15c(c+3)^2(c-3)}
      \partial^2 ({\bf J} {\bar{\bf \Phi }}^{(\frac{3}{2})})
      \right) \nonumber\\
      && +  \bar{C}_{(3)(2)}^{(2)} \left( \frac{72(8c^2+53c+57)}{
        5(c^2+26c+9)(2c+9)(c-3)}D^{i} {\bf J}D^{i}{\bf J}
      {\bar{\bf \Phi }}^{(2)} \right. \nonumber\\
      && - \frac{6(2c^3+7c^2+63c+90)}{5(c^2+26c+9)(2c+9)(c-3)}
      D^{i}{\bf J} D^{3-i} {\bar{\bf \Phi }}^{(2)}\nonumber\\
      && -  \frac{12c(4c^2+9c-99)}{5(c^2+26c+9)(2c+9)(c-3)}D^{3-0}
      {\bf J} {\bar{\bf \Phi }}^{(2)}
      - \frac{9(6c^3+17c^2-105c-36)}{10(c^2+26c+9)(2c+9)(c-3)}
      \partial^2 {\bar{\bf \Phi }}^{(2)}  \nonumber\\
      && + \frac{54(6c^2+21c-1)}{5(c^2+26c+9)(2c+9)(c-3)}{\bf J} D^{i}
      {\bf J} D^{i} {\bar{\bf \Phi }}^{(2)} \nonumber \\
      && \left. - \frac{3(14c^3+49c^2-129c-180)}{5(c^2+26c+9)(2c+9)(c-3)}
      D^{3-i}{\bf J} D^{i} {\bar{\bf \Phi }}^{(2)}
       +  \frac{72(4c^2+9c-99)}{5(c^2+26c+9)(2c+9)(c-3)}
      \partial {\bf J} {\bf J}
               {\bar{\bf \Phi }}^{(2)} \right) \nonumber\\
      && +  \bar{C}_{(\frac{5}{2})(2)}^{(\frac{5}{2})} \left(
      \frac{4 i c}{(c+15)(c-1)}\delta ^{i\alpha }
      (T^j)^{\alpha \beta }D^{i}{\bf J} D^{j}
      {\bar{\bf \Phi }}^{(\frac{5}{2}),\beta} \right.
      + \frac{120}{(c+15)(c-1)}\delta ^{i\alpha }
      {\bf J}D^{i} {\bf J} {\bar{\bf \Phi }}^{(\frac{5}{2}),\alpha}
      \nonumber\\
      && +  \left. \delta ^{i\alpha }\frac{12c}{(c+15)(c-1)}
      ( -D^{3-i}{\bf J}{\bar{\bf \Phi }}^{(\frac{5}{2}),\alpha}-
      \frac{1}{6} \partial D^{i}{\bar{\bf \Phi }}^{(\frac{5}{2}),\alpha}
      ) \right) \nonumber\\
      && + \bar{C}_{(2)(2)}^{(3)} \left( -\frac{3c(c-2)}{2(c+3)(7c-6)}
      \delta ^{i\alpha }D^{3-i}{\bar{\bf \Phi }}^{(3),\alpha} \right.
      \nonumber \\
      && -  \left. \frac{9(c-2)}{(c+3)(7c-6)}\delta ^{i\alpha }
      {\bf J}D^{i} {\bar{\bf \Phi }}^{(3),\alpha}   -
      \frac{24(4c-3)}{(c+3)(7c-6)}\delta ^{i\alpha }D^{i}
      {\bf J}{\bar{\bf \Phi }}^{(3),\alpha} \right) \nonumber \\
      && +   \frac{4}{9} \bar{C}_{(\frac{5}{2})(2)}^{(\frac{7}{2})}
      \delta ^{i\alpha }D^{i}{\bar{\bf \Phi }}^{(\frac{7}{2}),\alpha}+
      \bar{C}_{(3)(2)}^{(4)}{\bar{\bf \Phi }}^{(4)} \Bigg](Z_2)
    +   \cdots.
      \label{n3final}
\eea
It is straightforward to obtain all the component OPEs from
(\ref{n3final}) by applying the above superderivatives to both sides
and putting the fermionic coordinates to zero (\ref{comptofull}).
The previous nine pole terms in (\ref{ope1}), (\ref{ope2}),
(\ref{ope3}) and (\ref{ope4}) are distributed in (\ref{n3final}). 
Compared with the result in \cite{AK1607}, the presence of
the $SO(3)$ generators arise nontrivially.
Under the large $c$ limit, most of the terms in (\ref{n3final})
survive and this classical algebra should provide the asymptotic
symmetry algebra of the matrix extension of $AdS_3$ bulk theory
\cite{CHR1406,HR1503,CH1506}.

%%%%%%%%%%%%%%%%%%%%%%%%%%%%%%%%%%%%%%%%%%%%%%%%%%%%%%%%%%%%%%%%%%%%%
%%%%%%%%%%%%%%%%%%%%%%%%%%%%%%%%%%%%%%%%%%%%%%%%%%%%%%%%%%%%%%%%%%%%%%
\section{ Conclusions and outlook}
%9%%%%%%%%%%%%%%%%%%%%%%%%%%%%%%%%%%%%%%%%%%%%%%%%%%%%%%%%%%%%%%%%%%%%%
%%%%%%%%%%%%%%%%%%%%%%%%%%%%%%%%%%%%%%%%%%%%%%%%%%%%%%%%%%%%%%%%%%%%%

We have obtained the nontrivial ${\cal N}=3$ supersymmetric
OPE described in (\ref{n3final}) and
the right hand side of this OPE contain the various ${\cal N}=3$
multiplets summarized by (\ref{list}) where we have the explicit
forms for the coset fields with fixed $(N,M)$.

Although the vacuum character in the ${\cal N}=3$ Kazama-Suzuki model
will be found explicitly for generic $N$ and $M$, 
it will not be easy to 
observe the extra higher spin currents completely 
as we add them in the right hand side of
(\ref{n3final}). In order to determine the structure constants appearing 
in these extra higher spin currents, we should use the other Jacobi 
identities between them.
If the extra higher spin currents are present in the right hand side of
(\ref{n3final}) in the large $(N,M)$ values, we expect 
that they should appear 
linearly. 
The structure constants in them should contain the factors 
$(c-6)$ corresponding to $(N,M)=(2,2)$ case, 
$(c-9)$ corresponding to $(N,M)=(3,2)$ case and so on.
We may try to calculate the OPEs  manually
explicitly with arbitrary 
$(N,M)$ dependence
but this is beyond the scope of this paper.

The immediate question is that in order to see
the full structure of this algebra we should compute the
OPEs between the two lowest ${\cal N}=3$ multiplets,
the one in \cite{AK1607} and the one in this paper.
In other words, we should calculate the OPEs
$ {\bf \Phi}^{\left(\frac{3}{2}\right)}(Z_1)
\bar{\bf \Phi}^{(\frac{3}{2})}(Z_2)$ and $
\bar{\bf \Phi}^{\left(\frac{3}{2}\right)}(Z_1)
    \bar{\bf \Phi}^{(\frac{3}{2})}(Z_2)$ further.
    We expect that the ${\cal N}=3$ multiplets found in \cite{AK1607}
    will appear in the right hand sides of above OPEs.

    We can apply the fermionic construction of this paper to
    other coset model. If we do not have the $SO(2M N)_1$ and $SU(M)_{
      k +N}$ in the (\ref{KScoset}) with some change in the levels of
    the remaining factors, then we have the bosonic coset model.
    Recently, some of the nontrivial OPE in this model has been found
    in \cite{Ahn2011}. It would be interesting to study
    whether we can construct the ${\cal N}=1$ supersymmetric
    coset model at the level $k= N+M$.

    According to the classification of \cite{KS1,KS2},
    there is an orthogonal type of coset model. It is an open problem
    to describe whether we can observe its supersymmetric
    enhancement by taking the critical level as we do in this paper.
    
\vspace{.7cm}

%%%%%%%%%%%%%%%%%%%%%%%%%%%%%%%%%%%%%%%%%%%%%%%%%%%%%%%%%%%%%%
%%%%%%%%%%%%%%%%%%%%%%%%%%%%%%%%%%%%%%%%%%%%%%%%%%%%%%%%%%%%%%%
\centerline{\bf Acknowledgments}
%%%%%%%%%%%%%%%%%%%%%%%%%%%%%%%%%%%%%%%%%%%%%%%%%%%%%%%%%%%%%%%
%%%%%%%%%%%%%%%%%%%%%%%%%%%%%%%%%%%%%%%%%%%%%%%%%%%%%%%%%%%%%%%

This work of CA and JH was supported by
the National Research Foundation of Korea(NRF) grant
funded by the Korea government(MSIT)(No. 2020R1F1A1066893).
MHK was supported by an appointment to the YST Program at the APCTP through the Science and Technology Promotion Fund and Lottery Fund of the Korean Government. 
MHK was also supported by the Korean Local Governments - Gyeongsangbuk-do Province and Pohang City.

\newpage

\appendix

\renewcommand{\theequation}{\Alph{section}\mbox{.}\arabic{equation}}

%%%%%%%%%%%%%%%%%%%%%%%%%%%%%%%%%%%%%%%%%%%%%%%%%%%%%%%%%%%%%%%%
%%%%%%%%%%%%%%%%%%%%%%%%%%%%%%%%%%%%%%%%%%%%%%%%%%%%%%%%%%%%%%%%%%%%%
\section{The ${\cal N}=2$ superspace description of
${\cal N}=3$ multiplets}
%%%%%AAA%%%%%%%%%%%%%%%%%%%%%%%%%%%%%%%%%%%%%%%%%%%%%%%%%%%%%%%%%%%%%%%%
%%%%%%%%%%%%%%%%%%%%%%%%%%%%%%%%%%%%%%%%%%%%%%%%%%%%%%%%%%%%

We present each ${\cal N}=3$ multiplet (\ref{list}) in terms of two
${\cal N}=2$ multiplets as follows \footnote{We have
  $\theta = \theta^1-i \theta^2$ and $\bar{\theta}=-\theta^1 - i \theta^2$. By combining the two ${\cal N}=2$ multiplet and the third fermionic coordinate
  $\theta^3$, we can reexpress them in terms of the ${\cal N}=3$ multiplet
  as in \cite{AK1607}.}:
\bea
  {\bar{\bf W}}^{\left( \frac{3}{2} \right)}(Z)&=&
  2\bar{\psi}^{\left(\frac{3}{2}\right)}(z)+\theta
  ( \bar{\phi}^{(2),1}+ i \bar{\phi}^{(2),2} )(z) +
  \bar{\theta} (  -\bar{\phi}^{(2),1}+ i \bar{\phi}^{(2),2} )(z)+
  \theta \bar{\theta}\bar{\psi}^{\left(\frac{5}{2}\right),3}(z),
  \nonumber \\
  {\bar{\bf W}}^{\left( 2 \right)}(Z)&= &
  2\bar{\phi}^{\left( 2 \right),3}(z)+\theta ( -\bar{\psi}^{\left(
    \frac{5}{2} \right),1}- i \bar{\psi}^{\left( \frac{5}{2} \right),2} )(z)
  + \bar{\theta} (  -\bar{\psi}^{\left( \frac{5}{2} \right),1}+
  i \bar{\psi}^{\left( \frac{5}{2} \right),2} )(z)+ \theta \bar{\theta}
  2\bar{\phi}^{\left( 3 \right)}(z),
  \nonumber \\
  {\bar{\bf W}}^{\left( 2' \right)}(Z)&=&
  2\bar{\psi}^{\left( 2 \right)}(z)+\theta ( \bar{\phi}^{\left(
    \frac{5}{2} \right),1}+ i \bar{\phi}^{\left( \frac{5}{2} \right),2} )(z)
  + \bar{\theta} (  -\bar{\phi}^{\left( \frac{5}{2} \right),1}+
  i \bar{\phi}^{\left( \frac{5}{2} \right),2} )(z)+ \theta
  \bar{\theta}\bar{\psi}^{\left( 3 \right),3}(z), \nonumber \\
  {\bar{\bf W}}^{\left( \frac{5}{2} \right)}(Z)&= &
  2\bar{\phi}^{\left( \frac{5}{2} \right),3}(z)+\theta ( -
  \bar{\psi}^{\left( 3 \right),1}- i \bar{\psi}^{\left( 3 \right),2} )(z) +
  \bar{\theta} (  -\bar{\psi}^{\left( 3 \right),1}+ i \bar{\psi}^{\left(
    3 \right),2} )(z)+ \theta \bar{\theta}
  2\bar{\phi}^{\left( \frac{7}{2} \right)}(z), \nonumber \\
  {\bar{\bf W}}^{\left( \frac{5}{2}' \right), \alpha}(Z)&=&
  2\bar{\psi}^{\left(\frac{5}{2}\right), \alpha}(z)+
  \theta ( \bar{\phi}^{(3),1, \alpha}+ i \bar{\phi}^{(3),2, \alpha} )(z) +
  \bar{\theta} (  -\bar{\phi}^{(3),1, \alpha}+ i
  \bar{\phi}^{(3),2, \alpha} )(z)\nonu \\
  & + &  \theta \bar{\theta}\bar{\psi}^{\left(\frac{7}{2}\right),3, \alpha}(z),
  \nonumber \\
  {\bar{\bf W}}^{\left( 3 \right), \alpha}(Z)&=&
  2\bar{\phi}^{\left( 3 \right),3, \alpha}(z)+\theta
  ( -\bar{\psi}^{\left( \frac{7}{2} \right),1, \alpha}-
  i \bar{\psi}^{\left( \frac{7}{2} \right),2, \alpha} )(z) +
  \bar{\theta} (  -\bar{\psi}^{\left( \frac{7}{2} \right),1, \alpha}+ i
  \bar{\psi}^{\left( \frac{7}{2} \right),2, \alpha} )(z)\nonu \\
  & +&
  \theta \bar{\theta} 2\bar{\phi}^{\left( 4 \right), \alpha}(z),
  \nonumber \\
         {\bar{\bf W}}^{\left( 3' \right), \alpha}(Z)&=&
         2\bar{\psi}^{\left( 3 \right), \alpha}(z)+\theta
         ( \bar{\phi}^{\left( \frac{7}{2} \right),1, \alpha}+ i
         \bar{\phi}^{\left( \frac{7}{2} \right),2, \alpha} )(z) +
         \bar{\theta} (  -\bar{\phi}^{\left( \frac{7}{2} \right),1, \alpha}+
         i \bar{\phi}^{\left( \frac{7}{2} \right),2, \alpha} )(z)\nonu \\
         & +&
         \theta \bar{\theta}\bar{\psi}^{\left( 4 \right),3, \alpha}(z),
         \nonumber \\
{\bar{\bf W}}^{\left( \frac{7}{2} \right), \alpha}(Z)&=&
2\bar{\phi}^{\left( \frac{7}{2} \right),3, \alpha}(z)+\theta .
( -\bar{\psi}^{\left( 4 \right),1, \alpha}- i
\bar{\psi}^{\left( 4 \right),2, \alpha} )(z) +
\bar{\theta} (  -\bar{\psi}^{\left( 4 \right),1, \alpha}+ i
\bar{\psi}^{\left( 4 \right),2, \alpha} )(z)\nonu \\
& + &
\theta \bar{\theta} 2\bar{\phi}^{\left( \frac{9}{2} \right), \alpha}(z),
\nonumber \\
  {\tilde{\bf W}}^{\left( 3' \right)}(Z)&= &
  2\tilde{\psi}^{\left( 3 \right)}(z)+\theta
  ( \tilde{\phi}^{\left( \frac{7}{2} \right),1}+ i
  \tilde{\phi}^{\left( \frac{7}{2} \right),2} )(z) +
  \tilde{\theta} (  -\bar{\phi}^{\left( \frac{7}{2} \right),1}+
  i \tilde{\phi}^{\left( \frac{7}{2} \right),2} )(z)+
  \theta \bar{\theta}\tilde{\psi}^{\left( 4 \right),3}(z), \nonumber \\
  {\tilde{\bf W}}^{\left( \frac{7}{2} \right)}(Z)&=&
  2\tilde{\phi}^{\left( \frac{7}{2} \right),3}(z)+\theta
  ( -\tilde{\psi}^{\left( 4 \right),1}- i \tilde{\psi}^{\left( 4 \right),2} )(z)
  + \tilde{\theta} (  -\bar{\psi}^{\left( 4 \right),1}+ i
  \tilde{\psi}^{\left( 4 \right),2} )(z)\nonu \\
  & + & \theta \bar{\theta} 2\tilde{\phi}^{\left( \frac{9}{2} \right)}(z),
  \nonumber \\
 {\bar{\bf W}}^{\left( \frac{7}{2}' \right), \alpha}(Z)&=&
 2\bar{\psi}^{\left(\frac{7}{2}\right), \alpha}(z)+\theta
 ( \bar{\phi}^{(4),1, \alpha}+ i \bar{\phi}^{(4),2, \alpha} )(z) +
 \bar{\theta} (  -\bar{\phi}^{(4),1, \alpha}+
 i \bar{\phi}^{(4),2, \alpha} )(z)\nonu \\
 & + & \theta \bar{\theta}\bar{\psi}^{\left(\frac{9}{2}\right),3, \alpha}(z),
 \nonumber \\
           {\bar{\bf W}}^{\left( 4 \right), \alpha}(Z)&=&
           2\bar{\phi}^{\left( 4 \right),3, \alpha}(z)+\theta
           ( -\bar{\psi}^{\left( \frac{9}{2} \right),1, \alpha}-
           i \bar{\psi}^{\left( \frac{9}{2} \right),2, \alpha} )(z)
           + \bar{\theta}
           (  -\bar{\psi}^{\left( \frac{9}{2} \right),1, \alpha}+
           i \bar{\psi}^{\left( \frac{9}{2} \right),2, \alpha} )(z)\nonu \\
           & + & \theta \bar{\theta} 2\bar{\phi}^{\left( 5 \right), \alpha}(z),
           \nonumber \\
                     {\bar{\bf W}}^{\left( 4' \right)}(Z)&=&
                     2\bar{\psi}^{\left( 4 \right)}(z)+\theta
                     ( \bar{\phi}^{\left( \frac{9}{2} \right),1}+
                     i \bar{\phi}^{\left( \frac{9}{2} \right),2} )(z) +
                     \bar{\theta} (  -\bar{\phi}^{\left( \frac{9}{2} \right),1}+
                     i \bar{\phi}^{\left( \frac{9}{2} \right),2} )(z)+
                     \theta \bar{\theta}\bar{\psi}^{\left( 5 \right),3}(z),
                     \nonumber \\
                     {\bar{\bf W}}^{\left( \frac{9}{2} \right)}(Z)&=&
                     2\bar{\phi}^{\left( \frac{9}{2} \right),3}(z)+\theta
                     ( -\bar{\psi}^{\left( 5 \right),1}-
                     i \bar{\psi}^{\left( 5 \right),2} )(z) +
                     \bar{\theta} (  -\bar{\psi}^{\left( 5 \right),1}+
                     i \bar{\psi}^{\left( 5 \right),2}
                     )(z)+ \theta \bar{\theta}
                     2\bar{\phi}^{\left( \frac{11}{2} \right)}(z).
                     \nonumber 
\eea
The various OPEs in ${\cal N}=2$ superspace can be obtained by using the
package of \cite{KT} from the component results.
Or we can express (\ref{n3final}) in terms of four OPEs in ${\cal N}=2$
superspace.

%%%%%%%%%%%%%%%%%%%%%%%%%%%%%%%%%%%%%%%%%%%%%%%%%%%%%%%%%%%%%%%%
%%%%%%%%%%%%%%%%%%%%%%%%%%%%%%%%%%%%%%%%%%%%%%%%%%%%%%%%%%%%%%%%%%%%%
\section{The ${\cal N}=3$ multiplets after the
  decoupling of the spin-$\frac{1}{2}$ current of
${\cal N}=3$ superconformal algebra
}
%%%%%AAA%%%%%%%%%%%%%%%%%%%%%%%%%%%%%%%%%%%%%%%%%%%%%%%%%%%%%%%%%%%%%%%%
%%%%%%%%%%%%%%%%%%%%%%%%%%%%%%%%%%%%%%%%%%%%%%%%%%%%%%%%%%%%

We can decouple the spin-$\frac{1}{2}$ current $\Psi$
of ${\cal N}=3$ superconformal algebra and by using the
following transformation of the currents we obtain the corresponding
algebra explicitly 
\bea
{\bar{\psi}}^{\left( \frac{3}{2} \right)}& \rightarrow &
\bar{\psi}^{\left( \frac{3}{2} \right)},
\qquad
{\bar{\phi}}^{\left( 2 \right),i} \rightarrow 
\bar{\phi}^{\left( 2 \right),i},
\nonumber \\
{\bar{\psi}}^{\left( \frac{5}{2} \right),i}& \rightarrow &
\bar{\psi}^{\left( \frac{5}{2} \right),i}-\frac{3}{c}\Psi
\bar{\phi}^{\left( 2 \right),i},
\qquad
{\bar{\phi}}^{\left( 3 \right)} \rightarrow 
\bar{\phi}^{\left( 3 \right)}+
\frac{3}{2c} \Psi \partial \bar{\psi}^{\left( \frac{3}{2} \right)}
-\frac{9}{2c}\partial \Psi \bar{\psi}^{\left( \frac{3}{2} \right)},
\nonumber \\ 
{\bar{\psi}}^{\left( 2 \right)}&\rightarrow&
\bar{\psi}^{\left( 2 \right)},
\qquad
{\bar{\phi}}^{\left( \frac{5}{2} \right),i} \rightarrow
\bar{\phi}^{\left( \frac{5}{2} \right),i},
\nonumber \\
{\bar{\psi}}^{\left( 3 \right),i}&\rightarrow&
\bar{\psi}^{\left( 3 \right),i}-
\frac{3}{c}\Psi \bar{\phi}^{\left( \frac{5}{2} \right),i},
\qquad
{\bar{\phi}}^{\left( \frac{7}{2} \right)} \rightarrow
\bar{\phi}^{\left( \frac{7}{2} \right)}+
\frac{3}{2c} \Psi \partial \bar{\psi}^{\left( 2 \right)}-
\frac{6}{c}\partial \Psi \bar{\psi}^{\left( 2 \right)},
\nonumber \\ 
{\bar{\psi}}^{\left( \frac{5}{2} \right), \alpha}&\rightarrow&
\bar{\psi}^{\left( \frac{5}{2} \right), \alpha},
\qquad
{\bar{\phi}}^{\left( 3 \right),i, \alpha=j} \rightarrow
\bar{\phi}^{\left( 3 \right),i, \alpha=j} -
\frac{3}{c} i \varepsilon^{ijk}\Phi
\bar{\psi}^{\left( \frac{5}{2} \right), \alpha=k},
\nonumber \\
{\bar{\psi}}^{\left( \frac{7}{2} \right),i, \alpha=j}&\rightarrow&
\bar{\psi}^{\left( \frac{7}{2} \right),i, \alpha=j}-
\frac{3}{c}\Psi \left( \bar{\phi}^{\left( 3 \right),i, \alpha=j}+
\bar{\phi}^{\left( 3 \right),j, \alpha=i}-\delta^{ij}\bar{\phi}^{\left(
  3 \right),k, \alpha=k} \right),
\nonumber \\
{\bar{\phi}}^{\left( 4 \right), \alpha=i}&\rightarrow&
\bar{\phi}^{\left( 4 \right), \alpha=i}+
\frac{3}{2c}i \varepsilon^{ijk} \Psi  \bar{\psi}^{\left( \frac{7}{2} \right),j,
  \alpha=k}(z)+\frac{3}{2c} \Psi \partial \bar{\psi}^{\left(
  \frac{5}{2} \right), \alpha=i}
- 
\frac{15}{2c}\partial \Psi \bar{\psi}^{\left( \frac{5}{2} \right), \alpha=i},
\nonumber \\
{\bar{\psi}}^{\left( 3 \right), \alpha}&\rightarrow&
\bar{\psi}^{\left( 3 \right), \alpha},
\qquad
{\bar{\phi}}^{\left( \frac{7}{2} \right),i, \alpha=j} \rightarrow
\bar{\phi}^{\left( \frac{7}{2} \right),i, \alpha=j} -
\frac{3}{c} i \varepsilon^{ijk}\Phi \bar{\psi}^{\left( 3 \right), \alpha=k},
\nonumber \\
{\bar{\psi}}^{\left( 4 \right),i, \alpha=j}&\rightarrow&
\bar{\psi}^{\left( 4 \right),i, \alpha=j}-
\frac{3}{c}\Psi \left( \bar{\phi}^{\left( \frac{7}{2} \right),i, \alpha=j}+
\bar{\phi}^{\left( \frac{7}{2} \right),j, \alpha=i}-
\delta^{ij}\bar{\phi}^{\left( \frac{7}{2} \right),k, \alpha=k} \right),
\nonumber \\
{\bar{\phi}}^{\left( \frac{9}{2} \right), \alpha=i}&\rightarrow&
\bar{\phi}^{\left( \frac{9}{2} \right), \alpha=i}+
\frac{3}{2c}i \varepsilon^{ijk} \Psi  \bar{\psi}^{\left( 4 \right),j,
  \alpha=k}+\frac{3}{2c} \Psi \partial \bar{\psi}^{\left( 3 \right), \alpha=i}
-\frac{9}{c}\partial \Psi \bar{\psi}^{\left( 3 \right), \alpha=i},
\nonumber \\ 
{\tilde{\psi}}^{\left( 3 \right)} &\rightarrow&
\tilde{\psi}^{\left( 3 \right)},
\qquad
{\tilde{\phi}}^{\left( \frac{7}{2} \right),i} \rightarrow
\tilde{\phi}^{\left( \frac{7}{2} \right),i},
\nonumber \\
{\tilde{\psi}}^{\left( 4 \right),i}&\rightarrow&
\tilde{\psi}^{\left( 4 \right),i}-
\frac{3}{c}\Psi \tilde{\phi}^{\left( \frac{7}{2} \right),i},
\qquad
{\tilde{\phi}}^{\left( \frac{9}{2} \right)} \rightarrow
\tilde{\phi}^{\left( \frac{9}{2} \right)}+
\frac{3}{2c} \Psi \partial \tilde{\psi}^{\left( 3 \right)}-
\frac{9}{c}\partial \Psi \tilde{\psi}^{\left( 3 \right)},
\nonumber \\ 
{\bar{\psi}}^{\left( \frac{7}{2} \right), \alpha}&\rightarrow&
\bar{\psi}^{\left( \frac{7}{2} \right), \alpha},
\qquad
{\bar{\phi}}^{\left( 4 \right),i, \alpha=j} \rightarrow
\bar{\phi}^{\left( 4 \right),i, \alpha=j} -
\frac{3}{c} i \varepsilon^{ijk}\Phi \bar{\psi}^{\left( \frac{7}{2} \right),
  \alpha=k},
\nonumber \\
{\bar{\psi}}^{\left( \frac{9}{2} \right),i, \alpha=j}&\rightarrow&
\bar{\psi}^{\left( \frac{9}{2} \right),i, \alpha=j}-
\frac{3}{c}\Psi \left( \bar{\phi}^{\left( 4 \right),i, \alpha=j}+
\bar{\phi}^{\left( 4 \right),j, \alpha=i}-\delta^{ij}
\bar{\phi}^{\left( 4 \right),k, \alpha=k} \right),
\nonumber \\
{\bar{\phi}}^{\left( 5 \right), \alpha=i}&\rightarrow&
\bar{\phi}^{\left( 5 \right), \alpha=i}+
\frac{3}{2c}i \varepsilon^{ijk} \Psi
\bar{\psi}^{\left( \frac{9}{2} \right),j, \alpha=k}+\frac{3}{2c} \Psi \partial \bar{\psi}^{\left( \frac{7}{2} \right), \alpha=i}
 - 
\frac{21}{2c}\partial \Psi \bar{\psi}^{\left( \frac{7}{2} \right), \alpha=i},
\nonumber \\
{\bar{\psi}}^{\left( 4 \right)}&\rightarrow&
\bar{\psi}^{\left( 4 \right)},
\qquad
{\bar{\phi}}^{\left( \frac{9}{2} \right),i} \rightarrow
\bar{\phi}^{\left( \frac{9}{2} \right),i},
\nonumber \\
{\bar{\psi}}^{\left( 5 \right),i}&\rightarrow&
\bar{\psi}^{\left( 5 \right),i}-
\frac{3}{c}\Psi \bar{\phi}^{\left( \frac{9}{2} \right),i},
\qquad
{\bar{\phi}}^{\left( \frac{9}{2} \right)}  \rightarrow
\bar{\phi}^{\left( \frac{9}{2} \right)}+
\frac{3}{2c} \Psi \partial \bar{\psi}^{\left( 4 \right)}(z)-\frac{12}{c}\partial
\Psi \bar{\psi}^{\left( 4 \right)}.
\nonu
\eea
We expect that the result of this paper with \cite{AK1607} will produce
the extension of the previous work in \cite{Knizhnik,Bershadsky}.

%%%%%%%%%%%%%%%%%%%%%%%%%%%%%%%%%%%%%%%%%%%%%%%%%%%%%%%%%%%%%%%%%%%%%%%%%%%
%%%%%%%%%%%%%%%%%%%%%%%%%%%%%%%%%%%%%%%%%%%%%%%%%%%%%%%%%%%%%%%%%%%%%%%%%%

\end{document}